\documentclass[11pt]{article}
\usepackage{rpmacros}
\usepackage[dvipsnames,svgnames,x11names,hyperref]{xcolor}
\usepackage{amsthm}
\usepackage{xfrac}
\usepackage[T1]{fontenc}
\usepackage{enumitem}
\usepackage{verbatim}
\usepackage[russian,english]{babel}
\allowdisplaybreaks

\usepackage{thmtools, thm-restate}
\usepackage[colorlinks,pagebackref]{hyperref}
\hypersetup{
  linkcolor=[rgb]{0,0,0.4},
  citecolor=[rgb]{0, 0.4, 0},
  urlcolor=[rgb]{0.6, 0, 0}
}
\usepackage{mathrsfs,bbm}
\usepackage{todonotes}
\usepackage{tikz}
\usetikzlibrary{decorations.pathreplacing,backgrounds}
\usepackage{multirow}
\usepackage{setspace}
\usepackage[ruled, linesnumbered, vlined]{algorithm2e}
\SetKwComment{Comment}{/* }{ */}
\SetKwFunction{Prune}{Prune}
\SetKwFunction{FastDESolver}{FastDESolver}
\SetKwFunction{FastFESolver}{FastFESolver}
\SetKwFunction{ShortVector}{ShortVector}
\SetKwFunction{ConstructBasis}{ConstructBasis}
\SetKwFunction{UMultListRecovery}{UMultListRecovery}
\let\oldnl\nl% Store \nl in \oldnl
\newcommand{\nonl}{\renewcommand{\nl}{\let\nl\oldnl}}% Remove line number for one line

\usepackage{amsmath}

\renewcommand{\epsilon}{\varepsilon}
\newcommand{\eps}{\varepsilon}
\renewcommand{\tilde}{\widetilde}

\newif\ifdraft
\draftfalse
\ifdraft
\newcommand{\RGnote}[1]{\textcolor{BrickRed}{\guillemotleft RG: #1 \guillemotright}}
\newcommand{\VGnote}[1]{\textcolor{Blue}{\guillemotleft VG: #1 \guillemotright}}

\else
\newcommand{\RGnote}[1]{}
\newcommand{\VGnote}[1]{}
\fi

\newcommand{\frs}{\mathsf{FRS}}
\newcommand{\umult}{\mathsf{UMULT}}

\newcommand{\ehref}[1]{\href{mailto:#1}{#1}}

\newcommand{\calC}{\mathcal{C}}

\newcommand{\prune}{\mathsf{PRUNE}}

\newcommand{\ignore}[1]{}

\usepackage[nameinlink,capitalise]{cleveref}

\newtheorem{theorem}{Theorem}[section]
\newtheorem{lemma}[theorem]{Lemma}

\newtheorem{problem}[theorem]{Problem}
\newtheorem{question}[theorem]{Question}
\newtheorem{conjecture}[theorem]{Conjecture}
\newtheorem{corollary}[theorem]{Corollary}

\newtheorem{proposition}[theorem]{Proposition}
\theoremstyle{definition}

\newtheorem{definition}{Definition}

\newtheorem{remark}{Remark}

\newcommand{\LIST}{\mathsf{LIST}}
\usepackage{nth}
\usepackage{intcalc}
\usepackage{etoolbox}
\usepackage{xstring}

\usepackage{ifpdf}
\ifpdf
\else
\usepackage[quadpoints=false]{hypdvips}
\fi

\newcommand{\ECCCurl}[2]{http://eccc.hpi-web.de/report/\ifnumcomp{#1}{>}{93}{19}{20}#1/#2/}

\newcommand{\shortECCC}[2]{\texttt{\href{http://eccc.hpi-web.de/report/\ifnumcomp{#1}{>}{93}{19}{20}#1/#2/}{eccc:TR#1-#2}}}

\newcommand{\parseECCC}[1]{% Takes a string of the form TRxx/xxx or
\StrSubstitute{#1}{TR}{}[\tmpstring]%
\IfSubStr{\tmpstring}{/}{ %assuming string is of the form TRxx/xxx
\StrBefore{\tmpstring}{/}[\ecccyear]%
\StrBehind{\tmpstring}{/}[\ecccreport]%
}{% assuming string is of the form TRxx-xxx
\StrBefore{\tmpstring}{-}[\ecccyear]%
\StrBehind{\tmpstring}{-}[\ecccreport]%
}%
\shortECCC{\ecccyear}{\ecccreport}}

\usepackage[margin=1.1in]{geometry}
\title{Structure Theorems (and Fast Algorithms) for \\ List Recovery of Subspace-Design Codes} 
\author{
{Rohan Goyal\thanks{Massachusetts Institute of Technology, Cambridge \ehref{rohan\_g@mit.edu}.}}
\and 
{Venkatesan Guruswami\thanks{University of California, Berkeley \ehref{venkatg@berkeley.edu}.}  }
}

\date{}

\begin{document}

\maketitle
\thispagestyle{empty}
\begin{abstract}
    List recovery of error-correcting codes has emerged as a fundamental notion with broad applications across coding theory and theoretical computer science. Folded Reed-Solomon (FRS) and univariate multiplicity codes are explicit constructions which can be efficiently list-recovered up to capacity, namely a fraction of errors approaching $1-R$ where $R$ is the code rate.

    Chen and Zhang and related works showed that folded Reed-Solomon codes and linear codes must have list sizes exponential in $1/\epsilon$ for list-recovering from an error-fraction $1-R-\epsilon$.  These results suggest that one cannot list-recover FRS codes in time that is also polynomial in $1/\epsilon$.
    In contrast to such limitations, we show, extending algorithmic advances of Ashvinkumar, Habib, and Srivastava for list decoding, that even if the lists in the case of list-recovery are large, they are highly structured. 
    In particular, we can output a compact description of a set of size only $\ell^{O((\log \ell)/\epsilon)}$ which contains the relevant list, while running in time only polynomial in $1/\epsilon$ (the previously known compact description due to Guruswami and Wang had size $\approx n^{\ell/\epsilon}$). We also improve on the state-of-the-art algorithmic results for the task of list-recovery. 
\end{abstract}
\newpage
\tableofcontents
\newpage

\section{Introduction}

List recovery of error-correcting codes is a generalization of list and unique decoding, where instead of receiving a single corrupted symbol for each coordinate, the decoder receives a set of possible candidates for each position. This is often used to model \textit{decoding with soft information} i.e. where there is some uncertainty about each symbol but we would still like to recover all close-by codewords which are within the received sets in most coordinates.

Initially explicitly defined and named in \cite{GI01} to construct efficiently decodable codes, the notion of list recovery has found a surprisingly diverse array of applications spanning constructions of condensers and extractors~\cite{SU02, GuruswamiUV2009, KT2022}, group testing~\cite{INR2010, Cheraghchi_2011}, compressed sensing \cite{NPR2011}, heavy hitters~\cite{LNNTHeavy2016,DW22}, cryptography~\cite{HLR21}, quantum advantage~\cite{YZ24, ChaillouxT2024, OPI25} and more.
The problem of list recovery is formalized as follows:

\textbf{List-Recovery.} An error-correcting code $\mathcal C\subseteq \Sigma^n$ is said to be $(\delta, \ell, L)$ list-recoverable if given arbitrary subsets $L_1, \cdots, L_n \subseteq \Sigma$ each of size at most $\ell$, there are at most $L$ codewords $c\in \mathcal C$ for which there exists a $y \in L_1\times \cdots\times L_n$ such that $c$ and $y$ differ on at most $\delta n$ coordinates.

The case when $\ell=1$ corresponds to \textbf{list-decoding} (up to a fraction $\delta$ of errors).

It is easily seen that any rate $R$ error-correcting code $\mathcal C$ is not $(1-R+\epsilon, \ell, q^{\epsilon n})$ list-recoverable, even for $\ell =1$. On the other hand,  a random code of rate $R$ over a sufficiently large alphabet $\Sigma= \Sigma(R,\eps,\ell)$ is $(1-R - \epsilon, \ell, O(\ell/\epsilon))$ list-recoverable with high probability \cite{resch2020list}.  

Thus $1-R$ is the \emph{capacity} of list decoding/list recovery, as a function of the rate. 

A fundamental question then is whether we can construct explicit codes which are list-recoverable up to distance $1-R-\epsilon$ with small lists. 

 Guruswami and Rudra~\cite{GuruswamiR2008} constructed explicit codes, namely folded Reed-Solomon (FRS) codes, which they showed to be efficiently list-recoverable up to distance $1-R-\epsilon$ with polynomially sized lists. This achieved the optimal trade-off between rate and decoding radius. However, the output list-size $L$ in this result was very large, growing as $n^{\Omega(\ell/\eps)}$.

 Guruswami and Wang~\cite{GuruswamiW2013} showed that while the output list is large, it is contained in a low-dimensional subspace. This enabled improvements to the list-size by pre-coding the FRS code into an appropriate subspace-evasive set \cite{GuruswamiW2013,dvir2011subspaceevasivesets}. Furthermore, they also led to list-size improvements for FRS codes by \emph{pruning} the concerned subspace to only include codewords close to the input.
\iffalse
However, constructing, or even showing the existence of linear/additive codes, matching the output list size $L$ guaranteed for random codes has remained open since. 

It has since remained an open question to construct explicit codes with stronger guarantees matching those of the \textit{capacity} theorem. 
\fi

It has recently become apparent that the fundamental property of FRS codes (and related codes such as univariate multiplicity codes) which implies both the containment of the list in a low-dimensional subspace and an effective pruning strategy to bound its size is a certain \emph{subspace-design} property possessed by these codes.
Informally, a rate $R$ code $\mathcal C \subseteq (\FF^s)^n$ is a good subspace-design code if for every low-dimensional subspace $A$ of $\mathcal C$, the average dimension of $A \cap {\mathcal C}_i$ for random $i \in [n]$ drops to  $\approx R \cdot \dim(A)$, where ${\mathcal C}_i = \{ c \in \mathcal C \mid c_i = 0\}$.

The subspace-design property has since been used to bring down list sizes $L$ for $(1-R-\eps,\ell,L)$-list recovery of FRS codes to $O_{\ell, \epsilon}(1)$, i.e., independent of the code length $n$, in \cite{KoppartyRSW2023, Tamo2024}. Similar list size bounds have also been shown for random linear and randomly punctured Reed-Solomon codes \cite{LMS-focs25, li_et_al:LIPIcs.APPROX/RANDOM.2025.53}.  

The list-sizes $L$ achieved by these list recovery results are \emph{exponential} in $1/\eps$ (unlike the case of list-decoding where the list sizes can be polynomial, and in fact linear, in $1/\eps$~\cite{AGGLZ25,Srivastava2025,ChenZ2025}). Perhaps surprisingly, this has been shown to be inherent. Specifically, a FRS code, or any linear code, cannot be $(1-R-\epsilon, \ell, \ell^{o(R/\epsilon)})$ list-recoverable \cite{ChenZ2025, LMS-focs25, li_et_al:LIPIcs.APPROX/RANDOM.2025.53}. 
For a recent survey of exciting developments concerning list recovery, we refer to Resch and Venkitesh~\cite{RV25}. 

The above list-size lower bound for list recovery implies that it is not feasible to output the entire list in time $\poly(\ell/\epsilon, n)$.  It might, however, still be possible that the list, despite being large, has a compact description and such a description can be found in time $\poly(n, \ell/\epsilon)$. This is driving motivation for our work and we obtain results that make progress in this direction.

\subsection{Our Results}

Following \cite{GuruswamiW2013}, which showed the important structural result that not only is the list for the list-recovery problem for FRS codes up to distance $1-R-\epsilon$ polynomially bounded, but it is contained in an affine space of dimension $\frac{\ell}{\epsilon}$,   
 list-recovery results for $\frs$ codes have proceeded in two steps:
\begin{itemize}
\itemsep=0ex
    \item Finding the low-dimensional subspace containing all codewords.
    \item Bounding the number of close-by codewords in any low-dimensional space, either directly or using the subspace-design property, via suitable \textit{pruning} of the subspace.
\end{itemize}

The above two steps also govern associated algorithmic results for list recovery. Recently, \cite{GoyalHKS2025} showed that the first step can be implemented in near-linear time in the block length and polynomial in $1/\epsilon, \ell$.  The pruning step, as implemented in \cite{KoppartyRSW2023, Tamo2024}, takes time $\tilde O(n\cdot \exp(\tilde O((\log^2 \ell)/\epsilon)))$, which is exponential in $1/\eps$. For the case of list-decoding, a beautiful recent work~\cite{AHS25} showed that the pruning step can be implemented in $\tilde O(n\cdot \text{poly}(1/\eps))$ time.

Our work provides improvements on two fronts. First, we extend the \cite{AHS25} pruning algorithm to the task of list-recovery and improve on the best known algorithmic list-recovery bounds. We efficiently (in runtime $\tilde O(n\cdot \mathsf{poly}(L))$ where $L$ is the output list-size) find a list of size  $\poly(\ell/\epsilon)\cdot \ell^{O((\log \ell)/\epsilon)}$ containing the close-by codewords, which is slightly better than the earlier $(\ell/\epsilon)^{O((\log \ell)/\epsilon)}$ bound. We note that our list-size bound is worse than the near-optimal combinatorial bound of $(\ell/(R+\epsilon))^{(R+\epsilon)/\epsilon}$ shown recently in \cite{BCZ25} (see~\Cref{sec:related-work}) but improves on the best known algorithmic results.

Second, and more importantly, as a result complementing the list-size lower bound constructions in \cite{ChenZ2025}, we show that the lists at distance $1-R-\epsilon$ for subspace-design codes are contained in $\poly(\ell/\epsilon)$ many $(O((\log \ell)/\epsilon), \ell)$ \emph{sum-sets}. 
In particular, this gives a way to succinctly describe a set of size $\poly(\ell/\epsilon)\cdot \ell^{O((\log \ell)/\epsilon)}$ that contains all close-by codewords.  

\begin{theorem}
    For any good subspace-design code, given an $r$-dimensional linear space, and a list recovery input $L_1, \cdots, L_n$ with $|L_i|=\ell$, we can output $\poly(\log \ell, 1/\epsilon, r)$ many $ (O(\frac{R(\log (r\epsilon/R)+1)}{\epsilon}), \ell) $ \textit{sumsets} of codewords which contain all codewords from the linear space that are within $1- R -\epsilon$ distance of the list-recovery input in time $\tilde O(n\cdot \poly(\ell/\epsilon, r))$.
    
    A $(u, v)$ sumset is any set of the form $A_1+A_2+\cdots +A_u = \{a_1+a_2+\cdots+a_u\mid a_i \in A_i\}$ where $|A_i|\le v$.
\end{theorem}

Combining with the \cite{GoyalHKS2025} result, which lets us find the $r$-dimensional space efficiently for $\frs$ codes, we get that we can output $\poly(\log \ell, 1/\epsilon, r)$ many $ (O(\frac{R(\log (r\epsilon/R)+1)}{\epsilon}), \ell) $ \textit{sumsets} of codewords which contain all codewords that are within $1- R -\epsilon$ distance of the list-recovery input in time $\tilde O(n\cdot \poly(\ell/\epsilon))$. Thus, we get an $\tilde O(n\cdot \poly(\ell/\epsilon))$ algorithm to succinctly describe an $\tilde O(\ell^{(\log \ell)/\epsilon})$ sized set containing all the close-by codewords to the product set $L_1 \times L_2 \cdots \times L_n$.

\subsection{Overview of Techniques}

In a companion paper \cite{GGproximity25}, we showed that the subspace \textit{pruning} algorithm of \cite{AHS25} for the task of list-decoding can be made \emph{oblivious} to the received word, and only depend on the space being pruned. This was then leveraged to carry out the pruning using shared randomness to find, in one go, many codewords that all have agreements on a small common set of coordinates. 

This idea was then further developed to establish optimal proximity gaps for subspace-design codes.

Here, we again begin with the observation that running \cite{AHS25} with shared randomness for the task of list-recovery succeeds in finding each codeword within the affine space which is close to the input with high probability. At the end of running \cite{AHS25}, we are left with a \emph{small} set of coordinates where these successfully found codewords must all have agreements; the number of such codewords can then also be argued to be small. Combining the success probability and size of the lists found already gives our list-size bounds. 

A high probability bound and union bound over all close-by codewords lets us conclude that running algorithm only $\poly(\ell/\epsilon)$ times suffices to efficiently find all close-by codewords. 

We also prove a generic linear-algebraic result showing that the set of codewords completely agreeing with $\ell$ sized lists on $t$ coordinates is contained in a $(t, \ell)$ sum-set and the description of this sum-set can be found efficiently.

There are two major things we need to be careful about here:
\begin{itemize}
    \item The high probability and union bound argument can only work if the probability of finding any codeword is sufficiently high. Thus, a pruning algorithm which does not find any given codeword with probability at least $\poly(\frac{\epsilon}{\ell})$ would not suffice for bounding the number of sum-sets required even if it gets better list-size bounds. 
    In particular, it is crucial to use the codeword oblivious pruning algorithm. Else, we would be stuck in regimes where the number of sumsets required is quasi-polynomial in $\ell$.

    \item Directly implementing the \cite{AHS25} pruning algorithm would not guarantee that the number of coordinates all codewords in a single iteration agree on is \textit{small} (which in turn is necessary to ensure that the dimension $t$ of the sum-set is small). For this purpose, we need to ensure that our pruning algorithm is more \emph{aggressive}. In particular, we deviate from the previous pruning approaches and only fix coordinates which decrease the dimension of the current space by a $(1-\Omega(\epsilon))$ factor. This variant of the pruning is able to improve sum-set dimension from $\ell/\epsilon$ to $(\log \ell)/\epsilon$. 
\end{itemize}

\subsection{Related Work}
\label{sec:related-work}
In recent work, \cite{BCZ25} made a rather remarkable connection between subspace-designs and discrete Brascamp-Lieb inequalities using which they established near-optimal combinatorial list recovery bounds for subspace-design codes such as folded Reed-Solomon and univariate multiplicity codes (see Appendix~\ref{sec: Appendix List Recovery BCDZ}). In a companion work~\cite{brakensiek2025random},  they showed, adapting the techniques of \cite{LMS-focs25}, that their near-optimal bounds can be transferred from subspace-design codes to random linear and random Reed-Solomon codes as well.

It remains an interesting open question whether our bounds can be further improved to match theirs. We address their results again in \Cref{sec: Open and conjectures} and Appendix \ref{sec: Appendix List Recovery BCDZ}.

\section{Preliminaries}

We begin by introducing the basic coding-theoretic definitions we will be using.

\begin{definition}[Fractional Hamming Distance]
    For any two vectors $x, y$ in $\Sigma^n$ where $\Sigma$ is some alphabet, we define $\Delta(x, y) = \frac{|\{i\in [n]: x_i \ne y_i\}|}{n}$ to be the fraction of coordinates where they differ.
    
    For  a set $S\subseteq \Sigma^n$, we define $\Delta(x, S) = \min_{y\in S}\frac{|\{i\in [n]: x_i \ne y_i\}|}{n}$
    to be the minimum fractional Hamming distance of $x$ to its closest vector in $S$.
\end{definition}
Two fundamental quantities associated with a code are its rate and distance.

\begin{definition}[Distance of a code]
    For  code $\mathcal C \subseteq \Sigma^n$, we define its (relative) distance as $\Delta(\mathcal C) = \min_{x, y \in \mathcal C, x\ne y} \Delta(x,y)$
\end{definition}

\begin{definition}[Rate of a code]
    For a code $\mathcal C\subseteq \Sigma^n$, its rate $R(\mathcal C)$ is defined as 
    $R(\mathcal C)= \frac{\log_{|\Sigma|}|\mathcal C|}{n}$
\end{definition}

We will focus on linear/additive codes in this paper, defined as follows.

\begin{definition}[Additive codes]
Let $\FF$ be a finite field and let $\Sigma = \FF^s$ for some positive integer $s$. 
    A code $\mathcal C \subseteq \Sigma^n$ is said to be $\FF$-additive (or just additive when the field $\FF$ is clear from context or not relevant to the discussion) if $\mathcal C$ is an $\FF$-linear subspace of $\Sigma^n$. When $s=1$, the code is simply called a linear code.
\end{definition}

\subsection{The list-recovery problem}

This paper is focused on the problem of list-recovery, formally defined below.

\begin{problem}[List Recovery]
    For a code $\calC\subseteq \Sigma^n$, the \emph{list-recovery} problem with error rate $\delta$ and input list size $\ell$ is defined as follows. The input consists of  subsets $L_1, \cdots, L_n \subseteq \Sigma$ with $|L_i|=\ell$ for each $i$, and the task is to find the list of codewords 
    \[ \LIST(\calC, \delta, L_1\times L_2\times \cdots \times L_n)  := \{ c \in \calC \mid \Delta(c, L_1\times L_2\times \cdots \times L_n) \le \delta \} \ . \]

\end{problem}
The goal is to construct codes for which the list of codewords is small for all list-recovery inputs, subject to inherent trade-offs between rate, $\delta$ and $\ell$. The following shows an inherent limitation of linear and $\frs$ codes in this regard. We note that the trade-off $\delta \to 1-R$ is the best possible even for list-decoding.

\begin{theorem}[\cite{ChenZ2025, 
LMS-focs25,
li_et_al:LIPIcs.APPROX/RANDOM.2025.53}] 
    For all $R \in (0,1)$, $\eps \in (0,1-R)$, positive integers $\ell$, and finite fields $\FF$ with $|\FF|\ge \ell$, the following holds for every $\FF$-linear code $\calC \subset \Sigma^n$ of rate $R$ and sufficiently large block length. There exist inputs $L_1, \ldots, L_n \subset \Sigma$ with $|L_i|=\ell$ to the list-recovery problem for the code $\calC$ such that \[|\LIST(\calC, 1-R-\eps, L_1\times L_2\times \cdots \times L_n)  |  \ge \ell^{\lfloor R/\epsilon\rfloor}\ . \]
\end{theorem}
This result tells us that for constant rate codes, we cannot have algorithms with runtime polynomial in $1/\epsilon$ which output the entire list of codewords since the valid list itself could be significantly larger.

\subsection{Folded Reed-Solomon Codes}

The first codes that were shown to be list-recoverable up to capacity were folded Reed-Solomon codes in the work of \cite{GuruswamiR2008} and had the added property that they were list-recoverable algorithmically as well in polynomial time. In the following subsection, we formally define folded Reed-Solomon codes and present previous results about their list-decoding and recovery.
\begin{definition}[Folded Reed-Solomon Codes \cite{GuruswamiR2008}]\label{def: FRS Codes}
        Let $n, k, s$ be positive integers and $\FF_q$ be a field with $q>sn$ and $\gamma$ be an element of the field.
        Additionally, let $\alpha_1, \alpha_2, \ldots, \alpha_n \in \FF_q$ be distinct elements such that $\alpha_i \gamma^t \ne \alpha_j$ for all $i\ne j$ and $t<s$.
        
        The $s$-folded Reed-Solomon Code $\mathcal C\subseteq (\FF_q^s)^n$ with message length $k$ on $(\alpha_1, \ldots, \alpha_n)$ is given by: \[\frs_{q, n, k, s}( \bar\alpha)=\left\{\left(\begin{bmatrix}
            f(\alpha_1) \\
            f(\alpha_1\gamma)\\
            \vdots\\
            f(\alpha_1 \gamma^{s-1})
        \end{bmatrix}, \begin{bmatrix}
            f(\alpha_2) \\
            f(\alpha_2\gamma)\\
            \vdots\\
            f(\alpha_2 \gamma^{s-1})
        \end{bmatrix}, \cdots, \begin{bmatrix}
            f(\alpha_n) \\
            f(\alpha_n\gamma)\\
            \vdots\\
            f(\alpha_n \gamma^{s-1})
        \end{bmatrix} \right)\Bigg| f\in \FF_{q}[x], \deg f<k \right\}\]
\end{definition}

As mentioned previously, $\frs$ codes are extremely useful for the problem of list-recovery and we know this due to the following results which we now state informally:

\begin{theorem}[\cite{GuruswamiW2013, GuruswamiR2008, Kopparty2014}] \label{thm: Guruswami-Wang}
    For every choice of $\ell\in \mathbb N, \epsilon>0, R>0$, there exists an $s_0 = \Omega(\ell/\epsilon^2)$, such that any rate $R$ $s$-folded-Reed Solomon code $\cal C$ for $s>s_0$ contains an affine space $\cal H$ of dimension $r<O(\ell/\epsilon)$ such that $\LIST(\calC, 1-R-\epsilon, L_1, \ldots, L_n)\subseteq \mathcal H$.
\end{theorem}
Recent advances have shown that the task of finding the subspace as above can be achieved in near-linear time!

\begin{theorem}[\cite{GoyalHKS2025}]
    The affine space $\mathcal H$ from \Cref{thm: Guruswami-Wang} for list recovery of $\frs_{q, n, Rn, s}$ can be found in time $\tilde O(n\cdot \log q \cdot \poly(\ell/\epsilon, s))$ where $\tilde O$ ignores $\polylog n, \polylog\log q$ terms. 
\end{theorem}

\noindent More formal versions of the above results are stated in \Cref{thm: GuruswamiW2013}.

$\frs$ codes have been extremely useful in understanding the problems of list-recovery and list-decoding but the list-sizes in these results remained rather large with \Cref{thm: Guruswami-Wang} giving us a list size of $q^{O(\ell/\epsilon)}$ when list-recovering from a distance of $1-R-\epsilon$ and $\ell$ sized lists in each coordinate.

In a beautiful work, \cite{KoppartyRSW2023} proved a dramatic improvement in list-size bounds for list-decoding and list-recovery of $\frs$ codes proving list-size bounds of $O_{\epsilon, \ell}(1)$ i.e., constant when $\epsilon$ and $\ell$ are treated as constants. They also gave an efficient pruning algorithm which when combined with $\cite{GuruswamiW2013, GoyalHKS2025}$ gives an efficient algorithm to list-recover the entire list of codewords when $\ell, 1/\epsilon$ are treated as constants. The bounds and algorithms were improved and simplified in the follow up work of $\cite{Tamo2024}$. 

\begin{theorem}[\cite{KoppartyRSW2023, Tamo2024}]
    For every choice of $\ell\in \mathbb N, \epsilon>0, R>0$, there exists an $s_0 = \Omega(\ell/\epsilon^2)$, such that any rate $R$ $s$-folded-Reed Solomon code $\cal C$ for $s>s_0$ and lists $L_1, \cdots, L_n$ with $|L_i|=\ell$, we have $|\LIST( \mathcal C, 1-R-\eps,  L_1\times \cdots \times L_n)|\le (\ell/\epsilon)^{O(\frac{1+\log\ell}\epsilon)}.$

    Additionally, there exists an efficient randomized algorithm $\mathsf{PRUNE}$ that takes in an affine space $\mathcal H$ of dimension $r < \ell/\epsilon$ and outputs \[ \LIST(\calC, 1-R-\epsilon, L_1\times \ldots \times L_n) \cap \mathcal H\] in time $\tilde O(n\log q\cdot (\ell/\epsilon)^{O(\frac{1+\log\ell}\epsilon)}\cdot \poly(s))$.
\end{theorem}

One interesting property about \cite{KoppartyRSW2023, Tamo2024} was that for the task of list-decoding i.e. if $\ell=1$ then their algorithms worked for all linear codes given a low dimensional affine space. Thus, it remained open whether list bounds could be improved if the list-size bounds could be improved if one used any property of folded Reed-Solomon codes. 

In a line of exciting works \cite{Srivastava2025, AHS25} proved that there are algorithms that output lists of size only $O(1/\epsilon^2)$ when $\ell =1$. \cite{ChenZ2025} proved that combinatorially it can be shown that folded Reed-Solomon codes only have list sizes of $O(1/\epsilon)$. 

For us, we build on the work of \cite{AHS25}, which we state here:

\begin{theorem}[\cite{AHS25}]
    For every choice of $\epsilon>0, R>0$, there exists an $s_0 = \Omega(1/\epsilon^2)$, such that for any $s\ge s_0$ and any rate $R$ $s$-folded Reed Solomon code $\cal C$, and received word $y \in \FF^s$, it outputs a list $L$ with $|L|\le 1/\epsilon^2$ such that:
    \[L=\LIST(\mathcal C, 1-R-\epsilon, y)\]
with high probability and runs in time $\tilde O(n\cdot \log q \cdot  \poly(s, \tfrac{1}{\epsilon}))$. 
\end{theorem}

\subsection{Subspace-Design Codes}\label{subsec: Subspace design codes}

Subspace designs were introduced in~\cite{GX-stoc13} as a way to pre-code certain Reed-Solomon and algebraic-geometric codes so that they could then be list-decoded with small lists. Informally, a subspace design consists of a collection of subspaces $\mathcal H_i$ of some ambient space such that every low-dimensional space has non-trivial intersection with few of them. While~\cite{GX-stoc13} used random constructions of subspace designs, explicit constructions of subspace designs were given in~\cite{GK16}. Curiously, the construction was based on list-decodable codes such as folded RS and univariate multiplicity codes. Thus, subspace-designs were defined with a coding-theoretic motivation, and then themselves constructed using algebraic codes. Perhaps this was an early indication that they are natively a coding-theoretic concept, and recent results have cemented their fundamental role in coding theory~\cite{Srivastava2025, ChenZ2025,AHS25,brakensiek2025random,BCZ25,GGproximity25}.

We present the definition of subspace-design codes in the terminology of \cite{GGproximity25}. A similar concept was also defined in~\cite{ChenZ2025} where they called them subspace-designable codes.

\begin{definition}[Subspace-Design Property]\label{def: Subspace-designs}
    For any function $\tau: \mathbb N\rightarrow \R_{\le 1}$, an $\FF_q$-additive code $\mathcal C\subseteq (\FF_q^s)^n$ is said to be a $\tau$-\textit{subspace design} code if for every $r\in \mathbb N$, and every $\F_q$-linear subspace $\mathcal A$ of $\mathcal C$ of dimension at most $r$, the following holds: 
    \begin{equation*} \frac{\sum\limits_{i=1}^n \dim \mathcal A_i}{n}\le \dim(\mathcal A)\cdot \tau(r) \end{equation*}
    where $\mathcal A_i = \{a\in \mathcal A\mid a_i=0\}$.
\end{definition}

\begin{proposition}\label{prop: tau is non-decreasing}
    We can assume that if an additive code is a $\tau$-subspace design code, then $\tau$ is a non-decreasing function without loss of generality.
\end{proposition}
\begin{proof}
    Note that the guarantee for $\tau(r)$ holds for all subspaces of dimension at most $r$. Thus, even if we define $\tau'(r)\leftarrow\min_{r' \ge r} \tau(r')$ then the code is also a $\tau'$-subspace design which is non-decreasing and the guarantees only improve.
\end{proof}

\begin{lemma}\label{lem: tau is atleast R}
    For any $\tau$-\textit{subspace-design} $\FF_q$-additive code of rate $R$, we must have $\tau(r)\ge R-\frac{1}{n}$ for all $r\in \mathbb N$. 
\end{lemma}
\begin{proof}
     We just need to prove the result for $r=1$ since we can just take $\mathcal A$ of dimension at $1$. Pick any non-zero codeword $c$ which is zero on at least $Rn-1$ coordinates. Such a codeword exists since the code is additive and there are $|\Sigma|^{Rn}$ different codewords. Now, define the $1$-dimensional subspace $\mathcal A = \{\alpha \cdot c\mid \alpha \in \FF_q\}$. Clearly, $\dim \mathcal A_i =1$ for at least $R-\frac{1}{n}$ fraction of coordinates and so $\tau(1) \ge R-1/n$. 
\end{proof}

The subspace design property has proven to be extremely useful in recent times and as is turns out is at the heart of the stronger list size-bounds shown for list-decoding of folded Reed-Solomon codes and univariate multiplicity codes established in \cite{Srivastava2025, ChenZ2025, AHS25}.

\begin{theorem}[\cite{GK16}]
    $s$-folded Reed-Solomon codes are $\tau$-subspace-design codes for $\tau(r)= \frac{sR}{s-r+1}$ for all $1\le r\le s$ and $\tau(r)=1$ otherwise.
\end{theorem}

Thus, it can be shown that almost all results proven previously can be shown for all subspace-design codes.

\smallskip
One key property about subspace design codes is that all close-by codewords to any input word lie in a low-dimensional space. Specifically, we have the following trade-off between decoding radius and dimension. The following can be thought of as extending \Cref{thm: Guruswami-Wang} to all subspace-design codes.

\begin{lemma}\label{lem: Lists are contained in small dimensional affine spaces}
    For every $\tau$-subspace-design additive code $\mathcal C\subseteq \Sigma^n$, every positive integer $r$, and all input lists $L_1, \cdots, L_n\in \Sigma^n$ with $|L_i|\le \ell$, 
    the set 
    \[ \LIST(\mathcal C, 1-\tau(r)-\frac{\ell}{r}, L_1\times L_t\times \cdots\times L_n)\] is contained within a linear space of dimension less than $r$.
\end{lemma}
   The proof of the following for $\ell=1$ is implicit in \cite{ChenZ2025} and a more general version is proved in \cite{GGproximity25}.
\begin{proof}
     If $\dim \mathcal C < r$ there is nothing to prove. Otherwise, consider an arbitrary set of $r$ linearly independent $c^{(1)}, \cdots, c^{(r)}$ codewords in $\mathcal C$. 

Define $\mathcal A=\mathsf{span}(c^{(1)}, \cdots, c^{(r)})$. For each $j \in [n]$, let $\mathcal A_j = \{c\in \mathcal A\mid c_j \in L_j\}$.
     We now observe that  \[n\left(r\cdot \tau(r)+\ell\right)\ge \sum_{j=1}^n \left(\dim \mathcal A_j \right) \ge\sum_{j=1}^n|\{i\in [r]\mid c^{(i)}_j=y_j\}| = n \cdot \sum_{i=1}^{r} (1- \Delta(y,c^{(i)})) \ , \]

     where the first inequality follows from the subspace design property of $\mathcal C$. By averaging, it follows that $\Delta(y,c^{(i)}) \ge 1-  \tfrac{1}{r} - \tau(r)$ for some $i \in [r]$. Thus, not all the codewords $c^{(i)}$, $i=1,2,\dots,r$, can satisfy $\Delta(c^{(i)}, y) < 1 -\tfrac{1}{r} - \tau(r)$. 
\end{proof}

\begin{remark}
    Observe that the above proofs naturally extend to the \textit{average-radius list-recovery} as well.
\end{remark}

In light of the above results, we can informally think of the list-recovery or list-decoding of $\tau$-subspace design codes as proceeding in two modular steps:
\begin{description}
    \item[Step 1:] Given the code $\mathcal C$ and lists $L_1,\cdots, L_n$, find a low dimensional affine space $\mathcal H$ which contains all codewords close to the list-recovery input product space.
    \item[Step 2:] Find all the codewords inside a low dimensional affine space which are actually close to the list-recovery input product space.
\end{description}

The focus of \cite{KoppartyRSW2023, Tamo2024, AHS25} has been on performing the second step of this algorithm efficiently. Our current paper also focuses on getting structural results and efficient algorithms for step $2$ for all subspace-design codes.

One would expect that Step $1$ should in general be hard algorithmically for list-decoding or list-recovering arbitrary subspace-design codes.  In contrast, we have explicit codes like $\frs$ and $\umult$ for which step $1$ can be done efficiently. As mentioned previously, this has been the focus of works such as \cite{GuruswamiW2013,GoyalHKS2024, GoyalHKS2025}.
 
\section{Structure Theorems and Algorithms}

Before beginning our algorithm, we will review the previous works.

\subsection{A review of recent subspace-pruning algorithms for list-decoding}

As a warmup and review, we consider the following two algorithms which were proposed for list-decoding of Folded Reed-Solomon and Multiplicity codes but in fact work more generally for any $\tau-$subspace design codes. Both the algorithms take in an input affine space $\mathcal H$ of dimension $r$ and \textit{prune} this affine space until they are left with a single codeword. The algorithms have the broad structure outlined in \Cref{alg: Pruning algorithms}.

\begin{algorithm}
\setstretch{1.1}
    \caption{$\prune_{\mathcal D}(\mathcal H, y)$: Pruning a linear space}
    \label{alg: Pruning algorithms}
    \textbf{Input}: An $\FF_q$-affine space $\cal H$, a received word $y\in \FF_q^n$. \\
    If $\dim \mathcal H = 0$, output the only element in $\mathcal H$.\\
    Else, sample $i\sim \mathcal D_{\mathcal H, y}$ where $\mathcal D_{\mathcal H, y}$ is a distribution on $[n]$ and return $\prune_{\mathcal D}(\mathcal H_i, y)$ where $\mathcal H_i = \{h\in \mathcal H\mid h_i =y_i\}$. 
\end{algorithm}

The key method of the pruning algorithm is now in the choice of this distribution $\mathcal D$. This is where the algorithms of \cite{KoppartyRSW2023, Tamo2024} differ from the algorithm of \cite{AHS25}. The former works  use a uniform weighting on $[n]$, whereas \cite{AHS25} uses a clever non-uniform distribution tailored to the guarantee offered by the subspace-design property which depends on the affine space being pruned. We begin with the analysis of the uniform pruning algorithm.

\begin{theorem}[\cite{KoppartyRSW2023, Tamo2024}]
    Let $\mathcal C\subseteq \Sigma^n$ be an additive code with relative distance $\Delta(\mathcal C)$, let $\mathcal H\subseteq C$ be an affine space of dimension $r$, and let received word $y\in \Sigma^n$. Then for any, $c\in \mathcal C\cap \mathcal H$ such that $\Delta(c, y)<\Delta(\mathcal C)-\epsilon$, the algorithm $\prune_{\mathcal U_n}(\mathcal H, y)$ outputs $c$ with probability at least $\epsilon^{r}$ where $\mathcal U_n$ is the uniform distribution on $[n]$.
\end{theorem}

\begin{proof}
    Consider any $c'\ne c$ in $\mathcal H$, then observe that \[\Pr_{i\sim \mathcal U_n}[c_i = y_i \wedge c_i' \ne y_i]\ge \epsilon\]
since $c_i$ and $c_i'$ differ in at least $\Delta(\mathcal C)$ fraction of positions $i$, but $c$ and $y$ have relative Hamming distance at $\Delta(\mathcal C)-\epsilon$. Therefore, since we pick such an $i$ with probability $\ge \epsilon$, $c$ remains in the resulting affine space by pruning and the dimension also reduces.

 Now let $p_{c, r, y}$ be the minimum probability of $\prune_{\mathcal U_n}$ outputting $c$ across all $\mathcal H\subseteq \mathcal C$ of dimension at most $r$ which contain $c$. From the above, we have \[p_{c, r, y}\ge \epsilon \cdot p_{c, r-1, y}\ge \cdots \ge \epsilon^r p_{c, 0, r}=\epsilon^r\]
    as desired.
\end{proof}

\cite{AHS25} exploit the subspace design property of $\frs$ and $\umult$ codes to get better results. We state their algorithmic guarantee in the more general setting of subspace-design codes below. Our presentation of the results also varies slightly from \cite{AHS25} but the crux remains the same. 

\begin{theorem}\label{thm: AHS Pruning}
    Let $\mathcal C$ be a $\tau$-subspace design additive code. Let $\mathcal H\subseteq C$ be an affine space of dimension $r$, and let received word $y\in \Sigma^n$. Then for any, $c\in \mathcal C\cap \mathcal H$ such that $\Delta(c, y)<1- \tau(r)-\epsilon$, the algorithm $\prune_{\mathcal D_\epsilon}(\mathcal H, y)$ outputs $c$ with probability at least $\frac{\epsilon}{r+\epsilon}$ where $\mathcal{D}_{\epsilon}(\mathcal H, y)$  samples coordinate $i$ with probability $p_i = \frac{w_i}{\sum_{j=1}^n w_j}$ where:
    \[ w_i = \begin{cases}
        0 & \mathcal H_i = \emptyset \\
        0 & \mathcal H_i = \mathcal H\\ 
        \dim \mathcal H_i + \epsilon & \text{otherwise}\\
    \end{cases}\]
    Here, as above, $\mathcal H_i =\{h\in \mathcal H \mid h_i = y_i\}$.
\end{theorem}
    \begin{proof}
        Before beginning, we observe that it's not possible that all $w_i$ are $0$ as that would contradict the subspace design property as the set of coordinates where $c_i = y_i$ is at least a $\tau(r)+\epsilon$ fraction.

       As with \cite{AHS25}, we prove the result by induction on $r$. 

       The result clearly holds when $r=0$. Now, assume that the result holds for all $r'<r$. Then, if we let $\mathsf{p}_{\mathcal H, c, y}$ be the probability that $\prune_{\mathcal D_\epsilon}(\mathcal H, y)$ outputs $c$, we have:
        \[\mathsf{p}_{\mathcal H, c, y} = \sum_{i: y_i = c_i} p_i\cdot \mathsf{p}_{\mathcal H_i, c, y} \stackrel{\text{induction}}{\ge} \sum_{i: y_i = c_i, \mathcal H_i \ne \mathcal H} \frac{w_i}{\sum_{j=1}^n w_j}\cdot \frac{\epsilon}{w_i}\ge \epsilon\cdot\sum_{i: y_i = c_i, \mathcal H_i \ne \mathcal H} \frac{1}{\sum_{j=1}^n w_j}\]

        Now, if $\alpha$ fraction of the $i$'s have the property that $\mathcal H_i = \mathcal H$, then we have 
        \[\mathsf{p}_{\mathcal H, c, y}\ge \epsilon\cdot\sum_{i: y_i = c_i, \mathcal H_i \ne \mathcal H} \frac{1}{\sum_{j=1}^n w_j} \ge \epsilon\cdot \frac{1-\Delta(c, y)-\alpha}{\frac{\sum_{j=1}^n \left(\dim \mathcal H_i+\epsilon\right)}{n}-(r+\epsilon)\alpha}\ge \epsilon \cdot \frac{\tau(r)+\epsilon - \alpha}{\frac{\sum_{j=1}^n \dim \mathcal{H}_i}{n}+\epsilon - \alpha(r+\epsilon)}\]
        Now, we use the subspace-design property (\cref{def: Subspace-designs}) to conclude \[\mathsf{p}_{\mathcal H, c, y}\ge\epsilon\cdot \frac{\tau(r)+\epsilon - \alpha}{r\cdot \tau(r)+\epsilon - \alpha(r+\epsilon)}\ge\epsilon\cdot \frac{\tau(r)+\epsilon - \alpha}{(r+\epsilon)\cdot \tau(r)+\epsilon\cdot (r+\epsilon) - \alpha(r+\epsilon)}= \frac{\epsilon}{r+\epsilon} \ , \] 
        which concludes the proof.
    \end{proof}

\subsection{Our pruning algorithm for list-recovery}

\subsubsection{Discussion on approaches}

The algorithm of \cite{AHS25} is a clever improvement over the methods of \cite{KoppartyRSW2023, Tamo2024} with better weighting. Now, one can try generalizing this approach to list-recovery. A natural first attempt is as follows:
\begin{itemize}
    \item We do not just sample $i\in [n]$ but also an element $j\in [\ell]$ where $\ell$ is the size of the lists in each coordinate.
    \item We restrict ourselves then to the affine subspace of $\mathcal H$ agreeing with the $i$'th coordinate's $j$'th list-element.
\end{itemize}

This approach is in fact taken in \cite{Tamo2024}. He argues that for generic $\FF_q$-linear codes, this approach yields a list-size bound  of at most $\left(\frac{\ell}{\epsilon}\right)^r$ for decoding radius $1-\tau(\ell/\epsilon)-\epsilon$.  For $\tau$-subspace design codes, one can get a list size of at most $\left(\frac{\ell}{\epsilon}\right)^{O(\frac{1+\log \ell}{\epsilon})}$ for the same decoding radius. 
Thus, it is natural to ask whether the cleverer sampling of \cite{AHS25} could buy us improvements in list-size bounds. Even if it does, we seem to be stuck with an exponential time since we have a list-size lower bound of $\ell^{\Omega\left(\frac{R}{\epsilon}\right)}$ by \cite{ChenZ2025, li_et_al:LIPIcs.APPROX/RANDOM.2025.53}. We make the following key observations:
\begin{itemize}
    \item The above pruning algorithms can be made to be \emph{independent} of the received codeword; let $\mathcal H$ always be a linear space and $\mathcal H_i = \{h\in \mathcal H\mid h_i =0\}$. At the end, if our algorithm pin a set of coordinates $S$, then we can output a codeword $c\in \mathcal H$ such that $c$ agrees with $y$ on all coordinates in $S$. By our pruning guarantee, there is at most one such codeword. We show that this algorithm also offers the same guarantees proved above for list-decoding. This observation was also exploited in \cite{GGproximity25} to get optimal proximity gaps for subspace-design codes.
    \item Since we are now only computing a set of coordinates to pin on, but always set the value on the coordinate to be $0$, we may get many codewords within the lists in one go.
    \item We can make the pruning in \cite{AHS25} much more aggressive, only pinning on coordinates $i$ in which the dimensions of the subspaces $\mathcal H_i$ are a fair bit smaller (in relation to the dimension of $\mathcal H$). This can ensure that we reducing to a zero-dimensional space much faster.
\end{itemize}

Combining these ideas, we improve the list size bounds to roughly $\bigl( \frac{r+\epsilon}{\epsilon}\bigr) \cdot \ell^{O(\frac{1+\log \ell}{\epsilon})}$. A more formal theorem is given in \Cref{cor: list-size}. 

Our actual algorithm will in fact produce  a succinct description of a list of size $\poly(\ell/\epsilon)\cdot \ell^{O(\frac{1+\log \ell}{\epsilon})}$ which contains $\LIST(\mathcal C, 1-R-\epsilon, L_1\times\cdots \times L_n)$. For a more formal result, check \Cref{cor: Union of products}.
\subsubsection{Our pruning algorithm}

\begin{definition}
    The weight function $\mathsf{wt}_\eta$ of a linear space $\cal H$ is given by $\mathsf{wt}_\eta(\mathcal H) = \dim(\mathcal H)+\eta$.
\end{definition}

\begin{proposition}[Subspace Design property for weights of spaces]\label{prop: subspace design for weights}
    Let $\calC \subseteq \Sigma^n$ be a $\tau$-subspace design $\F_q$-additive code and let $\eta>0$. Let $A\subseteq \calC$ be any $\F_q$-linear space  of dimension $r>0$. Now, if $\mathcal A_i = \{a\in \mathcal A\mid a_i=0\}$ then: \[\frac{\sum\limits_{i=1}^n \mathsf{wt}_\eta(A_i)}{n}\le \mathsf{wt_\eta(A)}\cdot (\tau(r)+\eta)\]
\end{proposition}

\begin{proof}
   Indeed, we have  \[\frac{\sum\limits_{i=1}^n \mathsf{wt_\eta}(A_i)}{n}=\frac{\sum\limits_{i=1}^n \dim (A_i)}{n}+\eta\le r\cdot  \tau(r) +\eta\le \mathsf{wt_\eta(\mathcal A)}\cdot ( \tau(r)+\eta)  \ . \]
    The first step is by definition, the second is by the subspace-design property (\Cref{def: Subspace-designs}) and the final step holds since $\mathsf{wt}_\eta(\mathcal A)\ge r\ge 1$. We even had a similar result used inside the proof of \Cref{thm: AHS Pruning} above.
\end{proof}

Algorithm~\ref{alg: AHS-style pruning} takes in an input as a linear space $\mathcal H \subseteq (\FF_q^s)^n$ and outputs a set $T\subseteq [n]$ such that $|T|\le \dim |\mathcal H|$.

\begin{algorithm}\label{alg: AHS-style pruning}
\setstretch{1.1}
    \caption{$\mathsf{FPRUNE}(\mathcal H, \eta, \eta')$: Pruning a linear space}
    \textbf{Input}: An $\FF_q$-linear space $\cal H$ and a real $\eta>0$. Let $\mathcal H_i = \{c\in \mathcal H\mid c_i =0\}$\\
    If $\mathcal H = \{0\}$. Output $\Phi$.
    Sample $i$ from $[n]$ with probability $p_i$ where \[p_i = \begin{cases} 
      0 & \mathsf{wt}_\eta (\mathcal H_i)> (1-\eta') \mathsf{wt}_\eta(\mathcal H)\\
      \frac{\mathsf{wt}_\eta(\mathcal H_i)}{\sum\limits_{i: \frac{\mathrm{wt}_\eta(\mathcal H_i)}{1-\eta'}\le \mathsf{wt}_\eta(\mathcal H)}\mathsf{wt}_\eta(\mathcal H_i)} & \text{otherwise}\\
   \end{cases}\]
   Return $\{i\}\cup \mathsf{FPRUNE}(\mathcal H_i, \eta, \eta')$.
\end{algorithm}

Towards analyzing the performance of Algorithm~\ref{alg: AHS-style pruning} we introduce the following potential function.
\begin{definition}
Let $\Sigma = \F_q^s$ for some finite field $\F_q$ and integer $s \ge 1$.
For any $\F_q$-linear space $\mathcal{H} \subseteq \Sigma^n$, a vector $c \in \Sigma^n$, lists $L_1, \cdots, L_n \subset \Sigma$ and a set $T \subseteq [n]$ with the property that $\mathcal H$ is $0$ on all coordinates in $T$, we  define the following function: \[f_{\eta, \eta'}(\mathcal H, c, T, L_1,\cdots, L_n)=\begin{cases}
    0 & c_i\notin L_i \text{ for some $i\in T$}\\
    \frac{(1-\eta')^{|T|}}{\mathrm{wt}_\eta(\mathcal H)} \ . 
    \end{cases} 
\]
\end{definition}

We now provide a slightly stronger version of the \Cref{thm: AHS Pruning} lemma ensuring dimension reduction in every step. We note a few changes: \begin{itemize}
    \item Our pruning algorithm outputs only a set of \textit{agreement coordinates} and this step is independent of the lists in each coordinate after receiving the space $\mathcal H$. Thus, it is possible that our algorithm fixes sets of coordinates where there is no $c\in \mathcal H$ that agrees on these coordinates with any word.
    \item We have a faster dimension reduction and the associated guarantee that dimensions are much smaller each time.
    \item We prove the result using a \textit{monovariant} instead of induction to emphasise the \textit{top-down} nature of the pruning algorithm and a quantity that improves in each step. In contrast, \cite{AHS25}'s proof is inductive. Our proof method also directly applies to the original \cite{AHS25} lemma. A version of this theorem with $\eta'=0$ appeared in \cite{GGproximity25}.
\end{itemize}

\begin{lemma}[Stronger main \cite{AHS25} lemma]
Let $\Sigma = \F_q^s$ for some finite field $\F_q$ and integer $s \ge 1$.
Let $r$ be a positive integer and let $\eta>0$. Suppose we are given a $\tau$-subspace-design code $\mathcal C\subseteq \Sigma^n$, input lists $L_1, \cdots, L_n\subseteq \Sigma$ (no restrictions on the sizes of these lists or what they look like), and a  $\F_q$-linear space $\mathcal H \subset \Sigma^n$ of dimension $r$, and a subset $T \subseteq [n]$ so that $\mathcal H$ is identically $0$ on all of $T$. 

Suppose 
we sample a coordinate $i \in [n]$ with probability $p_i$ where \[p_i = \begin{cases} 
      0 & \mathsf{wt}_\eta (\mathcal H_i)> (1-\eta') \mathsf{wt}_\eta(\mathcal H)\\
      \frac{\mathsf{wt}_\eta(\mathcal H_i)}{\sum\limits_{i: \frac{\mathrm{wt}_\eta(\mathcal H_i)}{1-\eta'}\le \mathsf{wt}_\eta(\mathcal H)}\mathsf{wt}_\eta(\mathcal H_i)} & \text{otherwise} \ . \\
   \end{cases}\] 
   Then, for any $c\in \mathcal H$ satisfying 
   \begin{equation}
   \label{eq:lr-dist}
    \Delta(c, L_1\times L_2\times \cdots \times L_n)<1-\frac{\tau(r)+\eta}{1-\eta'} \ , 
    \end{equation}
    the following monotonicity holds: \[\mathbb E[f_{\eta, \eta'}(\mathcal H_i, c, T\cup \{i\}, L_1,\cdots, L_n)]\ge f_{\eta, \eta'}(\mathcal H, c, T, L_1,\cdots, L_n)\ . \]
\end{lemma}
\begin{proof}
    We drop the $L_1, \cdots, L_n$ from the function definition for convenience since it is fixed across the proof anyway.
    
    Observe that if $f_{\eta, \eta'}(\mathcal H,c, T)=0$ then there is nothing to prove. So we assume that $f_{\eta, \eta'}(\mathcal H,c, T)\ne 0$. 
    Let 
    \[ w_i = \mathrm{wt}_\eta(\mathcal H_i), \quad w=\mathrm{wt}_\eta(\mathcal H), \text{  and   } W=\sum\limits_{j: w_j\le (1-\eta')w}w_j \ . \]
    Suppose that $\alpha n$ of the coordinates satisfy $\mathsf{wt}_\eta(\mathcal H_i)> (1-\eta')\mathsf{wt}_\eta(\mathcal H)$. By the $\tau$-subspace-design property of code, we have 
    \begin{equation}
    \label{eq:bound-on-W}
        W \le \bigl(\tau(r)+\eta-\alpha(1-\eta') \bigr) w n
    \end{equation}
    We have 
    \begin{align*}
        \mathbb E[f_{\eta, \eta'}(\mathcal H_i,c, T\cup \{i\})] & = \Bigl(\sum_{i: c_i \in L_i} p_i\cdot f_{\eta, \eta'}(\mathcal H, c,T)\cdot\frac{w}{w_i}\Bigr)(1-\eta') \nonumber \\
        & = (1-\eta') f_{\eta, \eta'}(\mathcal H, c,T)\cdot \sum_{i: c_i\in L_i, ~ w_i\le (1-\eta')w}\frac{w_i}{W}\cdot \frac{w}{w_i} \nonumber \\
        &= (1-\eta') f_{\eta, \eta'}(\mathcal H, c, T) \sum_{i: c_i\in L_i, ~ w_i\le (1-\eta')w} \frac{w}{W}\nonumber \\
         & \ge (1-\eta') f_{\eta, \eta'}(\mathcal H, c, T) \frac{w}{W}  \bigl( |\{i: c_i \in L_i\}|-\alpha n  \bigr) \ . \nonumber \\
        \end{align*}
        Now by hypothesis \eqref{eq:lr-dist} we have $\frac{|\{i: c_i \in L_i\}|}{n} \ge \frac{\tau(r)+\eta}{1-\eta'}$, and we also have the upper bound \eqref{eq:bound-on-W} on $W$. Therefore we can bound
        \begin{align*}
        \mathbb E[f_{\eta, \eta'}(\mathcal H_i,c, T\cup \{i\})] & \ge (1-\eta')\frac{\frac{\tau(r)+\eta}{1-\eta'}-\alpha}{\tau(r)+\eta-\alpha(1-\eta')}\cdot f_{\eta, \eta'}(\mathcal H, c, T)  \\
        & = f_{\eta, \eta'}(\mathcal H, c, T)  \ ,
        \end{align*}
        completing the proof.
        \end{proof}

\begin{lemma}
\label{lem:succ-prob}
Let $r$ be a positive integer and let $\eta > 0$. Suppose we are given $\tau$-subspace design code $\mathcal C\subseteq \Sigma^n$, lists $L_1, \cdots, L_n\subseteq \Sigma$ (no restrictions on the sizes of these lists or what they look like), and a linear space $\mathcal H$ of dimension $r$. Then for any fixed $c\in \mathcal H$ satisfying $\Delta(c, L_1\times L_2\times \cdots \times L_n)<1-\frac{\tau(r)+\eta}{1-\eta'}$, the following holds: if algorithm $\mathsf{FPRUNE}(\mathcal H, \eta, \eta')$ returns a (random) set $T$, then we have  \[\mathbb E[X_{c,T}\cdot (1-\eta')^{|T|}]\ge \frac{\eta}{r+\eta}\]
    where $X_{c, T}$ is the indicator random variable for whether $c_i \in L_i$ for all $i\in T$.
\end{lemma}
\begin{proof}
    At the end if the output of the algorithm is a set $T$, then we know that \[\mathbb E[f_{\eta, \eta'}(\{0\}, c, T)]\ge f_{\eta, \eta'}({\mathcal H}, c, \emptyset)=\frac{1}{r+\eta}\implies \mathbb E[X_{c,T}\cdot (1-\eta')^{|T|}]\ge \frac{\eta}{r+\eta} \ .  \qedhere \]
\end{proof}

Also, observe that for any values $\beta_1, \beta_2, \cdots, \beta_{|T|}$, if $\mathcal H$ outputs $i_1, \cdots, i_{|T|}$, there is at most $1$ element $c\in \mathcal H$ such that  $c_{i_1}=\beta_1, c_{i_2}=\beta_2$ and so on.
\begin{theorem}\label{cor: list-size}
    For any integer $r>0$ and real $\eta>0$ given a $\tau$-subspace design code $\mathcal C\subseteq (\Sigma)^n$, lists $L_1, \cdots, L_n\subseteq \Sigma$ and $|L_i|=\ell$, and linear space $\mathcal H$ of dimension $r$, we have
    \[\left|\left\{c\in \mathcal H\mid \Delta(c, L_1\times L_2\times \cdots \times L_n)<1-\frac{ \tau(r)+\eta}{1-\eta'}\right\}\right|\le \ell^{\min(r, \frac{1}{\eta'}\left(1+\log{(r\eta')}\right)} \left(\frac{r}{\eta}+1\right)\]
\end{theorem}
\begin{proof} 

    We run the algorithm $\mathsf{FPRUNE}(\mathcal H, \eta, \eta')$ to get a set $T$. Now, if we consider all $c\in \mathcal H$ such that for all $j\in [T]$, we have $c_j \in L_j$, then we get a list $\LIST$ of codewords of at most $\ell^{|T|}$ codewords.
    
Note that $|T|\le r$ since the dimension always reduces by at least $1$. Also, we have \[|T|\le \Big\lceil\frac{1}{\eta'}(1+\log{(r\eta')})\Big\rceil\] since in the first $\frac{1}{\eta'}\log{(r\eta')}$ steps, the dimension reduces to at most $1/\eta'$ as  \[r\cdot (1-\eta')^{\frac{1}{\eta'}\log{(r\eta')}}\le r\cdot \frac{1}{r\eta'}\le \frac{1}{\eta'} \ , \] and then after that still always reduces by at least $1$ whenever an element is added to $T$.

\smallskip
\noindent Now, by \Cref{lem:succ-prob}, for each $c$ such that $\Delta(c,  L_1\times L_2\times \cdots \times L_n)< 1-\tfrac{ \tau(r)+\eta}{1-\eta'}$, we have 
\[ \Pr[c\in \LIST]\ge \frac{\eta}{r+\eta} \ . \]
\noindent Thus summing over all relevant codewords, we can conclude that \[\ell^{\min(r, \frac{1}{\eta'}(1+\log{(r\eta'))})}\ge \mathbb E[|\LIST|]\ge \frac{\eta}{r+\eta} \cdot \left|\left\{c\in \mathcal H\mid \Delta(c, L_1\times L_2\times \cdots \times L_n)<1-\frac{ \tau(r)+\eta}{1-\eta'}\right\}\right| \ , \]
which gives us the desired list-size bound.
\end{proof}

\begin{corollary}\label{cor: Pruning success}
    For any integer $r>0$ and real $\eta>0$ given a $\tau$-subspace design code $\mathcal C\subseteq \Sigma^n$, lists $L_1, \cdots, L_n\subseteq \Sigma$ and $|L_i|=\ell$, and linear space $\mathcal H$ of dimension $r$, let $\LIST = \{c\in \mathcal H\mid \Delta(c, L_1\times L_2\times \cdots \times L_n)<1-\frac{ \tau(r)+\eta}{1-\eta'}\}$. 
    
    If we run $\mathsf{FPRUNE}(\mathcal H, \eta)$ a total of $t=\frac{r+\eta}{\eta}\cdot (r \log \ell+\log(r/\eta+1)+t')$ times, then we get sets $T_1, \cdots, T_t$ such that \[\Pr[\forall c \in \LIST~ \exists j \in [t] \text{ such that } \forall i \in T_j: c_i \in L_i]\ge 1- e^{-t'}\]
\end{corollary}

\begin{proof}
    Observe that for any fixed $c\in \LIST$, $\Pr[\nexists j \in [t]\text{ such that } \forall i \in T_j: c_i \in L_i]\le (1-\frac{\eta}{r+\eta})^t\le e^{-\frac{\eta}{r+\eta}\cdot t}\le e^{-(\log |\LIST| + t')}\le \frac{e^{-t'}}{\LIST}$, where we have used the upper bound on $|\LIST|$ proved in \Cref{cor: list-size}.

    \noindent Now, we just conclude by union bounding over all the codewords in $\LIST$.
\end{proof}

This result tells us that we can get an efficient algorithm that outputs a description of few sets such that any relevant list-recovery codeword would be in all the lists of at least one of the output sets.

We would like to move away from this description of a few sets to a more natural description. An example of a nicer description would be not in terms of agreement coordinates,  but rather inside a more explicit space of codewords. We now turn to providing such a description.

\subsection{Sum-set structure from complete agreement}

In the following subsection, we consider the sets output by the list recovery algorithm from above and show that the lists as described before have an additional \textit{sum-set} structure.

The results in this subsection are in fact more generic linear algebraic claims and holds for all $\FF_q$-additive codes. Its benefit for $\tau$-subspace design codes only holds when combined with $\Cref{cor: Pruning success}$.

Let's begin by defining what we mean by a \textit{sum-set}-structure.

\begin{definition}[Sum Sets]
    $P\subseteq \mathcal C$ is an $(u, v)$ \textit{sum-set} in $\mathcal C$ if it is described by sets $A_1, A_2, \cdots A_{u'}$ with $|A_i|\le v$, $u'\le u$ and $A_i \in \mathcal C$ such that $P=A_1+\cdots +A_{u'}=\{a_1+a_2+\ldots + a_{u'} \mid a_i \in A_i\}$.
\end{definition}
Our key theorem in this subsection is that there is a natural way of going between agreeing on a few coordinates to a \textit{sum-set} structure.

\begin{theorem}\label{alg: Convert to product sets}
    Let $\mathcal H\subseteq (\FF_q^s)^n$ be an $r$-dimensional $\FF_q$-linear space and let $\{t_1,\cdots, t_m\}=T\subseteq [n]$ and $m\le r$ be such that $\{c\in \mathcal H\mid c_t = 0  \ \forall t \in T\} = \{0\}$. Then the following holds
    for all subsets $L_{t_1}, L_{t_2}, \cdots, L_{t_m} \subseteq \FF_q^s$ with $|L_i|=\ell$: 
    
    There exist sets $A_{t_i} \subseteq \mathcal H$ with $|A_{t_i}|\le \ell$ for each $i\in [m]$ such that \[\{c\in \mathcal H: c_{t }\in L_{t} \ \forall t \in T\}\subseteq A_{t_1}+A_{t_2}+\cdots A_{t_m} = \{a_1+a_2+\ldots + a_m \mid a_i\in A_{t_{i}}\} \ . \] 
    In fact, there is an algorithm $\mathsf{REDUCE}$ which outputs $A_{t_1}, \cdots, A_{t_m}$ and runs in time $\tilde O(n\cdot\ell\cdot \log q \cdot \poly(s, r))$.  
\end{theorem}

\begin{proof}
    Observe that since $\{c\in \mathcal H\mid c_i = 0 \ \forall i \in T\} = \{0\}$, we have reduced $\mathcal H$ from dimension $r$ to dimension $0$. 
    
    We can now view $\left(\FF_q^s\right)^n$ as $\FF_q^{sn}$ and restrict our view to the $sm$ coordinates in $T$. Now, by performing Gaussian elimination, we can find $r$ positions within these $sm$ coordinates such that the values on them are linearly independent. We call these positions the \textit{independence} positions.

    Thus, we can let $T_1, \cdots, T_m$ be sets with each $T_i \subseteq [s]$ and $\sum |T_i|=r$ such that $\mathcal H$ restricted to $(T_1, \cdots, T_m)$ is full dimensional, i.e., the projection of $\mathcal H$ onto the coordinates determined by $T_1,\cdots, T_m$ has dimension $r$ equal to the dimension of $\mathcal H$ itself. 

    Now, if we restrict ourselves not to being in all the $m$ lists entirely - but only on being in the respective lists restricted to the coordinates of the $T_i$s, we can only get a larger list.
 
 We use this to create the set $A_{t_i}$. \[A_{t_i} = \{c\in \mathcal H \mid {c_{t_j}}_{\mid T_j}=0 \ \forall j\ne i, {c_{t_i}}_{\mid T_i}\in {L_{t_i}}_{\mid T_i}\}\]
    Since, for all $\beta \in L_i$, there is a unique $c\in \mathcal H$ such that $c\in \mathcal H \mid {c_{t_j}}_{\mid T_j}=0 \forall j\ne i, {c_{t_i}}_{\mid T_i}=\beta_{\mid T_i}$, we have that $|A_{t_i}|\le \ell$ for each $i$. 

    Now, also for each $\beta_1\in L_{t_1}, \beta_2 \in L_{t_2},\cdots, \beta_m \in L_{t_m}$, there is a unique element $a_{\beta_i}$ in $A_{t_i}$ such that ${a_{\beta_i}}_{\mid T_i} = {\beta_i}_{\mid T_i}$. Thus, $a_{\vec\beta}=a_{\beta_1}+\cdots + a_{\beta_m}$ is the unique element in $\mathcal H$ such that \[{a_{\vec\beta}}_{\mid T_i}={\beta_i}_{\mid T_i}\] for all $i\in [m]$. Thus, defining the sets $A_{t_i}$ as above works as desired. 

    This protocol is clearly algorithmic and once we restrict $\mathcal H$ to just the coordinates of $T$, Gaussian elimination to find \textit{independence} positions runs in time $\tilde O(\log q)\poly(r, s)$. Now, computing the sets $A_{t_i}$ requires solving an $r$-dimensional linear system and we have $\ell \cdot m$  such computations to do. Solving each of these linear systems takes $\tilde O(\log q \cdot \poly(r, s))$ time and the claimed result follows. 
\end{proof}

\subsection{Combining pruning and sum-set structure results}

\begin{theorem}\label{cor: Union of products}
    For any integer $r, s>0$ and real $\eta>0$ given a $\tau$-subspace design code $\mathcal C\subseteq (\FF_q^{s})^n$, and lists $L_1, \cdots, L_n\subseteq \FF_q^s$ and $|L_i|=\ell$, and a linear space $\mathcal H$ of dimension $r$, there is an algorithm running in time $\tilde{O}(n\log q\cdot \poly(\ell, s, r, \frac{1}{\eta\cdot \eta'}))$, which outputs the descriptions of $t$ many $(\min(r, \lceil\frac{1}{\eta'}(1+\log{(r\eta')}\rceil),\ell)$ sum-sets $P_1, \cdots, P_t$ such that:
    \[\Pr[\{c\in \mathcal H\mid \Delta(c, L_1\times L_2\times \cdots \times L_n)<1-\frac{ \tau(r)+\eta}{1-\eta'}\subseteq \bigcup_{j=1}^t P_j] \ge 1-e^{-t'}\]
    and runs in time $\tilde{O}(n\cdot \log q \cdot \poly(\ell, r, s, \frac{1}{\eta})\cdot t')$ where $t=\frac{r+\eta}{\eta}\cdot (r \log \ell+\log(r/\eta+1)+t')$
\end{theorem}

\begin{proof}
    Run Algorithm~\ref{alg: AHS-style pruning}, $\mathsf{FPRUNE}(\mathcal H, \eta, \eta')$ a total of  $t$ times and for each of output sets $T$, invoke the algorithm $\mathsf{REDUCE}$ from \Cref{alg: Convert to product sets}. The success probability follows from \Cref{cor: Pruning success}. 
\end{proof}

\section{Efficient list-recovery of $\frs$ codes}

As we discussed earlier, the current approaches for list recovering or list decoding subspace-design codes can be broken into two modular steps:
\begin{itemize}
    \item \textbf{Step 1:} Find a low dimensional affine space in which all close codewords lie.
    \item \textbf{Step 2:} Prune this space to be left with only a few codewords.
\end{itemize}

In general, step $1$ might not be feasible for an arbitrary code, but due to \cref{cor: Union of products} and \cref{thm: AHS Pruning}, we know that step $2$ can always be done efficiently. Thus, a natural question concerns the codes for which one can implement step $1$ efficiently.

Indeed, our motivation to study these bounds was heavily inspired by $\frs$ and $\umult$ codes which indeed do have efficient algorithms to achieve step $1$.
\begin{theorem}[\cite{GuruswamiW2013, GoyalHKS2025}]\label{thm: GuruswamiW2013}
    There exists a deterministic algorithm such that for any rate $R$ $s$-folded Reed-Solomon code and input lists $L_1, \cdots, L_n\subseteq \FF_q^s$ with $|L_i|=\ell$, and parameter $r\le s$ outputs an affine space $\mathcal A$ with the following properties:
    \begin{itemize}
        \item $\dim \mathcal A \le r$
        \item 
        $\LIST(\mathcal C, 1-\frac{s}{s-r}\left(\frac{\ell}{r}+R\right), L_1\times L_2\times \cdots \times L_n) \subseteq \mathcal A$.
        \item The algorithm runs in time $\tilde O(n\cdot \log q\cdot \poly(\ell, s))$.
    \end{itemize}
\end{theorem}

The runtime $\tilde O(n\cdot \log q\cdot \poly(\ell, s))$ is due to \cite{GoyalHKS2025}
and the original \cite{GuruswamiW2013} has a runtime of $\poly(n,\log q,\ell, s)$.

\begin{theorem}
\label{thm:FRS-nearlinear}
    For every choice of $\ell\in \mathbb N, \epsilon>0, R>0$,  any rate $R$ $s$-folded-Reed Solomon code $\cal C$ for $s>s_0 :=  \frac{16 (R+\epsilon)\ell}{\epsilon^2}$ has the following property. 
    
    There exists an algorithm that takes in input lists $L_1, \cdots, L_n \subseteq \F_q^s$ of size $\ell$ each, runs in time $\tilde O(n\cdot \log q\cdot \poly(s))$, and outputs the descriptions of $t = \tilde O(\ell^2/\epsilon^3)$ many 
    \[ \left(\left\lceil \frac{2(R+\epsilon)(\log (2e\ell/(R+\epsilon)))}{\epsilon}\right\rceil, \ell\right) \text{ sum-sets } P_1, \cdots, P_t \]
    such that 
        $\LIST(\mathcal C, 1-R-\eps , L_1\times L_2\times \cdots \times L_n) \subseteq  \bigcup_{i=1}^t P_i$ with probability $\ge \frac{1}{2}$. 
\end{theorem}

\begin{proof}
 We apply \Cref{thm: GuruswamiW2013} with parameters $r=\frac{4\ell}{\epsilon}$, $s_0=\frac{16(R+\epsilon)\ell}{\epsilon^2}$.  It asserts that there is an affine space $\mathcal A$ of dimension at most $r$ containing  \[ \LIST(\mathcal C, 1-\tfrac{s}{s-r}\bigl(\tfrac{\ell}{r}+R\bigr), L_1\times L_2\times \cdots \times L_n) \ . \] 
    By our choices of $s_0, r$, we get that $\frac{\ell}{r}=\frac{\epsilon}{4}$ and \[\frac{s}{s-r}\le 1+\frac{r\left(\frac{4(R+\epsilon)}{\epsilon}\right)}{r\left(\frac{4(R+\epsilon)}{\epsilon}-1\right)}\le \frac{4R+4\epsilon}{4R+3\epsilon}\le \frac{4R+2\epsilon}{4R+\epsilon}=\frac{R+\epsilon/2}{R+\epsilon/4}\]
    Thus, $\frac{s}{s-r}(\tfrac{\ell}{r}+R)\le R+\epsilon/2$. In particular, 
    \begin{equation}
      \label{eq:frs-sumset1}  
    \LIST(\mathcal C, 1-R-\eps , L_1\times L_2\times \cdots \times L_n) \subseteq \mathcal A \ , 
    \end{equation}
    where $\mathcal A$ is at most $\frac{4\ell}{\epsilon}$ dimensional. This affine space $\mathcal A$ is also output by the \cite{GoyalHKS2025} algorithm. 

    Now, we can apply \Cref{cor: Union of products} with parameters $t'=1$, $\eta = \frac{\epsilon}{4}$, $\eta'= \frac{\epsilon}{2(R+\epsilon)}$. 
    This algorithm outputs $P_1, \cdots, P_t$ which are all $\left(\left\lceil\frac{(\log (er\eta'))}{\eta'}\right\rceil, \ell\right)$ \textit{sum-sets} such that 
    \begin{equation}
        \label{eq:frs-sumset2}
    \Pr[\{c\in \mathcal A\mid \Delta(c, L_1\times L_2\times \cdots \times L_n)<1-\frac{ \tau(r)+\eta}{1-\eta'}\} \subseteq \bigcup_{j=1}^t P_j] \ge 1-e^{-t'}\ge1/2 \ . 
    \end{equation}
    Note that \Cref{cor: Union of products} works with linear spaces and not affine space. Observe that this is easy to fix as we can translate the entire problem i.e. all the input lists and $\mathcal A$ by one of the elements of the affine space $\mathcal A$. This results in $\mathcal A$ becoming a subspace. This element can also be added back to each \textit{sum-set} outputted by the \Cref{cor: Union of products} algorithm to get \textit{sum-sets} that work for the original untranslated problem.

    Observe that again by our setting of parameters, we have that $\tau(r)\le R\cdot \frac{s}{s-r+1}<R\cdot \frac{s}{s-r}\le R+\epsilon /4$, $\eta \le \epsilon/4$, and thus 
    \begin{equation}
        \label{eq:frs-sumset3}
\frac{\tau(r)+\eta}{1-\eta'}< \frac{R+\epsilon/2}{(2R+\epsilon)/(2R+2\epsilon)}=R+\epsilon \ . 
    \end{equation}
Combining \eqref{eq:frs-sumset1}, \eqref{eq:frs-sumset2}, and \eqref{eq:frs-sumset3}, we conclude 
 \[\Pr[\LIST(\mathcal C, 1-R-\eps, L_1\times L_2\times \cdots \times L_n) \subseteq \bigcup_{j=1}^t P_j] \ge 1/2\ . \]
    Additionally, we have that \[\left\lceil\frac{(\log (er\eta'))}{\eta'}\right\rceil=\left\lceil\frac{2(R+\epsilon)\log (2e\ell/(R+\epsilon))}{\epsilon}\right\rceil\]
    as desired and $t = \lceil\frac{r+\eta}{\eta}\left(r\log \ell + \left(\log (r/\eta+1)\right) +1\right)\rceil = \tilde O(\ell^2/\epsilon^3)$. 
\end{proof}

We also get the following theorem:

\begin{theorem}
    For every choice of $\ell \in \mathbb N$, $\epsilon>0$, $R> 0$, $s_0 = \frac{16(R+\epsilon)\ell}{\epsilon^2}$, such that  for any rate $R$ $s$ folded Reed-Solomon code $\cal C$ with $s>s_0$, we have that \[\left|\LIST(\mathcal C, 1 -R-\eps, L_1\times L_2\times \cdots \times L_n)\right|\le O\left(\frac{\ell}{\epsilon^2}\right)\cdot \ell^{2(\ln (2e\ell/(R+\epsilon))\cdot \frac{R+\epsilon}{\epsilon}} \ . \]
\end{theorem}
\begin{proof}
    By combining the choice of parameters described above with \Cref{thm: GuruswamiW2013} and \Cref{cor: list-size}, we get that $|\LIST|\le \ell^{\frac{1}{\eta'}\ln (er\eta')}\cdot (r/\eta+1)$ where $r=4\ell/\epsilon$, $\eta=\epsilon/4$, $\epsilon/2(R+\epsilon)$.

    This gives us a bound of \[|\LIST|\le \left(16\ell/\epsilon^2+4/\epsilon\right)\cdot\ell^{\frac{2(R+\epsilon)}{\epsilon}\cdot \ln(2e\ell/(R+\epsilon))} \ . \qedhere \]
\end{proof}

%\iffalse
\section{Open Questions and Conjectures}\label{sec: Open and conjectures}

%\iffalse
In recent work, \cite{BCZ25}
prove stronger combinatorial list-recovery bounds by insightfully connecting the subspace-design property to Brascamp-Lieb inequalities. \footnote{We note that our results were obtained before \cite{BCZ25} was posted.}

Their list size bounds are significantly better than what we are able to achieve using our pruning techniques. Below we state their result in our terminology of subspace-design codes, and 
for completeness, in Appendix \ref{sec: Appendix List Recovery BCDZ}, we present their proof.

\begin{theorem}[\cite{BCZ25} List-Recovery Size Bounds]\label{thm: BCZ25 List Recovery bounds}
    If $\mathcal C\subseteq \Sigma^n$ is a $\tau$-subspace design code, then for any linear space $\mathcal H$ of dimension $r$, and subsets $L_1, \cdots, L_n \subseteq \Sigma$ with each $|L_i|=\ell$, and any $\eps > 0$, we have \[|\{c\in \mathcal H\mid \Delta(c, L_1\times L_2\times \cdots\times L_n)<1-\tau(r)-\epsilon\}|\le (\ell/(\tau(r)+\epsilon))^{\frac{\tau(r)+\epsilon}{\epsilon}} \ . \]
\end{theorem}

This raises the question whether we can use our algorithmic techniques in combination with their ideas to prune the lists further? Along these lines, we make the following conjecture.

\begin{conjecture}
    For every choice of $\ell\in \mathbb N, \epsilon>0, R>0$, there exists an $s_0 = \Omega(\ell/\epsilon^2)$, such that any rate $R$ $s$-folded Reed-Solomon code $\cal C$ for $s>s_0$, there exist $\poly(\ell/\epsilon)$ many $\left(O\left(\frac{R+\epsilon}{\epsilon}\right), \ell\right)$ sum-sets $P_1, \cdots, P_t$ such that \[\{c\in \mathcal C\mid \Delta(c, L_1\times L_2\times \cdots \times L_n)<1-R-\epsilon\}=\bigcup_{i=1}^t P_i.\]
\end{conjecture}
This conjecture would give a combinatorial list-recovery bound of $\poly(\ell/\epsilon)\cdot \ell^{O\left(\frac{R}{\epsilon}\right)}$ which we know to be optimal due to \cite{ChenZ2025} and \cite{li_et_al:LIPIcs.APPROX/RANDOM.2025.53}. Note that the above form of the conjecture is quite strong as it states that the list is exactly a union of a few low-dimensional sum-sets and not just contained within such a union.

It is also natural to make the same conjecture for random linear codes and random Reed-Solomon codes since we now know matching list recovery size bounds from \cite{LMS-focs25, brakensiek2025random, li_et_al:LIPIcs.APPROX/RANDOM.2025.53}.

Our algorithm is randomized and derandomizing it is another interesting direction.

\begin{question}(Informal)
    Can the above pruning algorithms be made deterministic while also running in $\poly(\ell, 1/\epsilon)$ time?
\end{question}

\cite{AHS25} show that the pruning for list-decoding can indeed be implemented with deterministic time complexity $\tilde O(n)\cdot (1/\epsilon)^{O(2^{1/\epsilon})}$. It would be interesting to improve this complexity or to even match it for list-recovery.

\section{Acknowledgements}

We thank Joshua Brakensiek for useful discussions and feedback.

V.G. is supported by a Simons Investigator award, NSF grant CCF-2211972, and ONR grant N00014-24-1-2491.

R.G. is supported by (Yael Tauman Kalai’s) grant from  Defense Advanced Research Projects Agency
(DARPA) under Contract No. HR0011-25-C-0300. Any opinions, findings and conclusions or recommendations expressed in this material are those of the author(s) and do not necessarily reflect the views of the Defense Advanced Research Projects Agency (DARPA). 

Part of this work was carried out while R.G. was visiting the Simons Institute for the Theory of Computing.

\addcontentsline{toc}{section}{References}
{\small \bibliographystyle{alpha}
   \bibliography{Structure-FRS-bib}
}
\appendix 
\section{List-recovery bounds via Brascamp-Lieb for subspace-design codes} 
\label{sec: Appendix List Recovery BCDZ}

The following results and proofs are entirely due to \cite{BCZ25} (see their Theorems 4.2 and 5.2)  and we state and reprove them here in our terminology of subspace-design codes.

The amazing insight in \cite{BCZ25} was to connect the subspace-design property to discrete Brascamp Lieb inequalities \cite{CCE09, CDKS13, CDKS24}. They also provide a self-contained proof of the version of the inequality needed for this application, as stated below. 

\begin{theorem}[Brascamp-Lieb]\label{thm: BL ineq}
    Let $V$ be a $d$-dimensional vector space over $\F_q$. For each $i\in [m]$, consider a scalar $s_i\ge 0$ and a linear map $\pi_i: V\rightarrow V_i$, where $V_i$ is a $d_i$-dimensional vector space over $\FF_q$. Assume further that for all subspaces $U\subseteq V$, the following holds: \[\dim U \le \sum_{i=1}^n s_i \dim(\pi_i(U)) \ . \]
    Then, for every probability distribution $X$ over $V$, we have \[H(X)\le \sum_{i=1}^n s_i H(\pi_i(X)) \ , \]
    where as usual $H(D)$ denotes the Shannon entropy of a distribution $D$. 
\end{theorem}

\begin{lemma}\label{lem: subspace-design for BL}
    If $\mathcal C$ is a $\tau$-subspace-design code, and $\pi_i$ is the projection map from $\mathcal C$ to the $i$th coordinate, then for very linear space $\mathcal H$ of dimension $r$, we have
    \begin{equation}
        \label{eq:dim-bound}
    \dim \mathcal H\le \frac{\sum\limits_{i=1}^n \dim(\pi_i(\mathcal H))}{(1-\tau(r))n} \  .
    \end{equation}
\end{lemma}
\begin{proof}
    Observe that \eqref{eq:dim-bound} is  equivalent to 
    \[
   \sum_{i=1}^n \left( \dim(\mathcal H)-\dim(\pi_i(\mathcal H))\right)\le \tau(r)\cdot r\cdot n \iff \sum_{i=1}^n \dim(\mathcal H_i)\le \tau(r)\cdot r\cdot n \ , \]
    where $\mathcal H_i = \{c  \in \mathcal H \mid c_i = 0\}$, which is precisely the definition of a subspace-design code.
\end{proof}

\begin{lemma}\label{lem: BL to list bounds}
    Let $\mathcal C$ be a $\tau$-subspace-design code, and $\pi_i$ the projection map from $\mathcal C$ to the $i$th coordinate. Then, the following holds for every linear space $\mathcal H$ of dimension $r$ and every distribution $X$ on $\mathcal H$: \[H(X)\le \frac{\sum\limits_{i=1}^n H(\pi_i(X))}{(1-\tau(r))n} \]
\end{lemma}
\begin{proof}
    Let $V=\mathcal H$ and $V_i = \pi_i(V)$. Observe, that for all $U\subseteq V$ of dimension $r_U\le r$, by \Cref{lem: subspace-design for BL}, we have \[\dim U \le \sum_{i=1}^n \frac{1}{(1-\tau(r_U))n}\dim(\pi_i(U))\le \sum_{i=1}^n \frac{1}{(1-\tau(r))n}\dim(\pi_i(U))\] since $\tau$ is non-decreasing without loss of generality by \Cref{prop: tau is non-decreasing}. 
The claim now follows by applying \Cref{thm: BL ineq} to $V$. 
\end{proof}

\begin{theorem}
    If $\mathcal C\subseteq \Sigma^n$ is a $\tau$-subspace design code, then for any linear space $\mathcal H \subset \mathcal C$ of dimension $r$, and input lists $L_1, \cdots, L_n$ with $L_i\subseteq \Sigma$ and $|L_i|=\ell$, we have \[|\{c\in \mathcal H\mid \Delta(c, L_1\times L_2\times \cdots\times L_n)<1-\tau(r)-\epsilon\}|\le \bigl(\tfrac{\ell}{\tau(r)+\epsilon}\bigr)^{(\tau(r)+\epsilon)/\epsilon} \  . \]
\end{theorem}
\begin{proof}
    Let $\LIST = \{c\in \mathcal H\mid \Delta(c, L_1\times L_2\times \cdots\times L_n)<1-\tau(r)-\epsilon\}$. Let $L = |\LIST|$ and $X$ be the uniform distribution on $\LIST$. Now, by \Cref{lem: BL to list bounds}, we have 
  \begin{equation}
  \label{eq:BL1}
      (1-\tau(r)) n \cdot \log L \le \sum\limits_{i=1}^n H(\pi_i(X)) \ . 
  \end{equation}
For $i \in [n]$, let $\rho_i = \frac{|\{c\in \LIST\mid c_i\in L_i\}|}{L}$, and denote $Y_i$ to be the indicator random variable for the event $\pi_i(X) \in L_i$. By the chain rule, 
\begin{align}
    H(\pi_i(X)) &  = H(Y_i) + H(\pi_i(X) | Y_i)  \nonumber 
\\ 
& \le  H(Y_i) + \rho_i \log \ell + (1-\rho_i) \log (L (1-\rho_i)) \nonumber \\
& = - \rho_i \log \rho_i  + \rho_i \log \ell + (1-\rho_i) \log L  \label{eq:BL2} \ . 
\end{align}
Define $\rho$ to be the average of the $\rho_i$'s. 
Combining \eqref{eq:BL1} and \eqref{eq:BL2}, we have
\begin{align}
    (\rho-\tau(r))\log L & \le \rho \log \ell - \frac{\sum_{i=1}^n \rho_i \log \rho_i}{n} \nonumber  \\ 
& \le \rho \log (\ell/\rho)  \label{eq:BL3}
\end{align}
where in the second step we used Jensen's inequality and concavity of the function $-x\log x$ for $x\in (0,1)$. 

By the definition of $\LIST$, we know $\rho \ge \tau(r)+\epsilon$. Using this with \eqref{eq:BL3}, we have the desired list-size upper bound $L\le \bigl(\frac{\ell}{\tau(r)+\epsilon}\bigr)^{\frac{\tau(r)+\epsilon}{\epsilon}}$.  
\end{proof}

\end{document}